\documentclass[floatfix,superscriptaddress,aps,amsmath,amssymb]{revtex4}

\usepackage[english]{babel}
\usepackage{graphicx}
\usepackage{color}
\usepackage{amssymb,amsmath}
\usepackage{xspace}
\usepackage{upgreek}
\usepackage{psfrag}
\usepackage{filemod}
\usepackage{siunitx}

\newcommand{\rg}{{\bf r}}

\newcommand{\qg}{{\bf q}}

\newcommand{\rhog}{\boldsymbol\rho}
\newcommand{\epseff }{\varepsilon_{\mathrm{eff}}}
\newcommand{\eps}{\varepsilon}

\newcommand{\kzeff}{k_z^{\mathrm{eff}}}
\newcommand{\Eref}{E_{\mathrm{ref}}}
\newcommand{\lsca}{\ell_s}
\newcommand{\lext}{\ell_e}
\newcommand{\lcor}{\ell_\varepsilon}

\begin{document}
	
\title{A model for full-field optical coherence tomography in scattering media}
	
\author{Ugo Tricoli}
\affiliation{ONERA, The French Aerospace Lab, Base A\'erienne 701,13661 Salon Cedex AIR, France}
\affiliation{Institut Langevin, ESPCI Paris, CNRS, PSL University, 1 rue Jussieu, 75005 Paris, France}
\author{R\'emi Carminati}
\altaffiliation{remi.carminati@espci.fr}
\affiliation{Institut Langevin, ESPCI Paris, CNRS, PSL University, 1 rue Jussieu, 75005 Paris, France}

\begin{abstract}
We develop a model of full-field optical coherence tomography (FF-OCT) that includes a description of partial temporal and spatial coherence, together with a mean-field scattering theory going beyond the Born approximation. Based on explicit expressions of the FF-OCT signal, we discuss essential features of FF-OCT imaging, such as the influence of partial coherence on the optical transfer function, and on the decay of the signal with depth. We derive the conditions under which the spatially averaged signal exhibits a pure exponential decay, providing a clear frame for the use of the Beer-Lambert law for quantitative measurements of the extinction length in scattering media. 
\end{abstract}

\maketitle

\section{Introduction}

Since its initial development~\cite{huang1991optical}, Optical Coherence Tomography (OCT) has proven its ability to image inside scattering materials with micrometer resolution in three-dimensions~\cite{schmitt1999optical}. The possibility to access cellular structures in tissues at millimeter depth has been a breakthrough in biomedical optics, which stimulated a fast and broad dissemination of the technique.

The OCT setup is essentially a low coherence Michelson interferometer, in which one arm collects the light backscattered from the sample, while the other arm produces a reference beam reflected on a mirror. The main feature of an OCT setup is the ability to decouple the depth (longitudinal) and transverse resolutions~\cite{fercher2003optical}. Depth resolution is produced by temporal coherence gating, and is controlled by the spectral width of the incident light. Transverse resolution is controlled by the numerical aperture (NA) of the microscope objective in the sample arm. In scanning OCT (S-OCT)~\cite{huang1991optical,nassif2004vivo,huber2006fourier,hoeling2000optical,zhang2005adaptive}, a point-by-point image is formed by three-dimensional scanning of a focal spot. OCT systems recording {\it en face} images in planes perpendicular to the optical axis have also been developed, using spatially coherent illumination as in wide-field OCT (WD-OCT) ~\cite{bourquin2001optical,bordenave2002wide,laubscher2002video}, or spatially incoherent illumination as in full-field OCT (FF-OCT) ~\cite{vabre2002thermal,fercher2000thermal}.

Since OCT is expected to collect the singly backscattered photons, the signal is substantially affected by multiple scattering, whose contribution overcomes the signal at depths larger than the scattering mean free path~\cite{schmitt1997model}. Several approaches have been followed to decrease the mutiple scattering contribution by reducing the weight of long light paths, including spatial filtering through confocal detection~\cite{webb1996confocal}, time gating \cite{hee1993femtosecond}, or polarization gating~\cite{schmitt1992use,macdonald2017numerical}. Other strategies address an inverse problem to correct {\it a posteriori} for multiple scattering, and increase resolution and penetration depth. Approaches based on interferometric synthetic aperture microscopy~\cite{marks2009partially}, or computational adaptive optics~\cite{adie2012computational}, have proven to be successful. Interestingly, it has been shown that the effect of multiple scattering depends on the degree of spatial coherence of the illuminating beam~\cite{marks2009partially}. It was also demonstrated that aberrations in the sample arm do not influence the transverse resolution in FF-OCT using spatially incoherent light~\cite{xiao2016full}. Recently, a method based on the measurement of the reflection matrix has demonstrated an efficient discrimination between singly and multiply scattered light, with an unprecedented increase in the OCT working depth~\cite{badon2016smart}, thus pushing the limits of optical microscopy in highly scattering media~\cite{badon2017multiple}.

Despite the success of OCT, from basic to clinical studies, a comprehensive theoretical model of the OCT signal, that handles a description of partial coherence together with a realistic scattering model (beyond the Born approximation), is still missing. In this paper we develop such a model, and use it to discuss different aspects of FF-OCT imaging. We study the influence of partial coherence on the optical transfer function. We also address the question of the decay of the signal with depth, that is captured by a mean-field scattering approach. Depending on the degree of spatial coherence and on the numerical aperture of the illumination/collection optics, we derive the conditions under which the spatially averaged FF-OCT signal exhibits a pure exponential decay with depth. The provides a clear frame for the use of the Beer-Lambert law for quantitative measurements of the extinction length in scattering media. 

\section{Scattering model}

\begin{figure}[h]
	\centering
	\includegraphics[width=11cm]{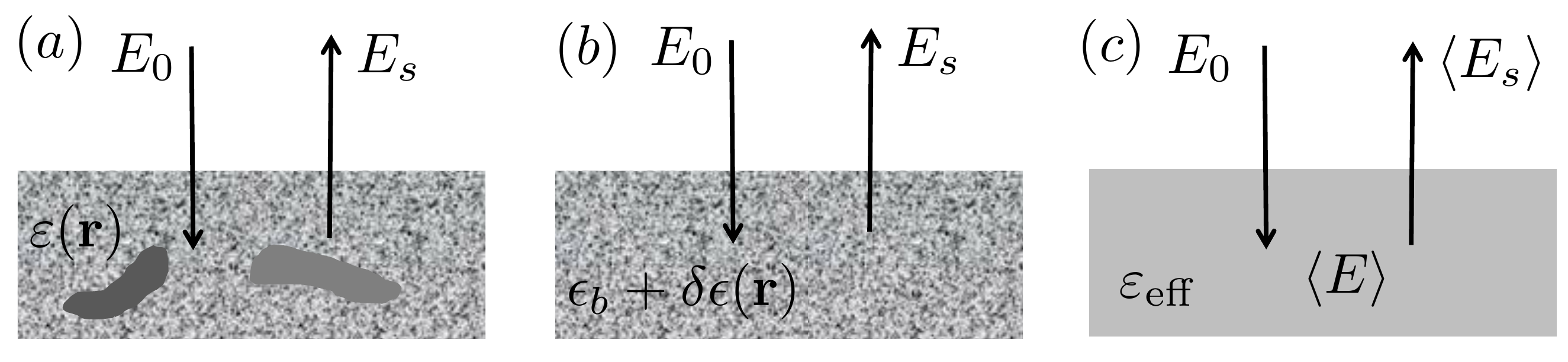}
	\caption{(a)~Heterogeneous medium with large scale inhomogeneities immersed in a scattering background. (b)~Purely scattering medium with small scale fluctuations of the dielectric function around a uniform value $\eps_b$. (c)~Effective medium characterizing the propagation of the average field.}
	\label{fig:sample}
\end{figure}

We consider a sample made of large-scale inhomogeneities immersed in a scattering medium made of randomly distributed scattering centers, as represented in Fig.~\ref{fig:sample}(a).The sample is characterized by a dielectric function $\eps(\rg)$ that we write as $\eps(\rg)=\eps_b + \eps_d(\rg) + \delta\eps(\rg)$, where $\eps_b$ denotes a uniform background, $\eps_d(\rg)$ is a large-scale deterministic dielectric function varying on a scale $L_\varepsilon\gtrsim \lambda$, with $\lambda$ the wavelength in vacuum, and $\delta\eps(\rg)$ is a real random variable describing a disordered distribution of small-scale scattering centers. We assume that $\delta\eps(\rg)$ satisfies $\langle \delta\eps(\rg) \rangle = 0$, the brackets denoting a statistical average over an ensemble of realizations of the disordered scattering background, and that the correlation function $\langle \delta\eps(\rg) \delta\eps(\rg^\prime) \rangle$ is of the form $f(|\rg-\rg^\prime|/\ell_\varepsilon)$, where $f$ is a positive decaying function with range close to unity. This defines $\ell_\varepsilon$ as the microscopic length scale of the disordered scattering medium, and we assume $\ell_\varepsilon\lesssim \lambda$. Note that this description is very general. For example by setting $\eps_d(\rg)=0$ we would describe a purely scattering medium, as that represented in Fig.~\ref{fig:sample}(b) (OCT is sometimes used to measure the extinction length $\lext$ in such materials). By setting $\delta\eps(\rg)=0$ we would describe large-scale objects in a uniform background.

In the statistical approach, in which the scattering background is treated as a random medium, the scattered field is a random variable. In a given realization of the sample (deterministic large-scale objects immersed in one statistical realization of the scattering background), the total field can be written $E=\langle E \rangle + \delta E$. In OCT, one measures the backscattered field  $E_s = E-E_0$, $E_0$ being the incident field, and we can also write $E_s = \langle E_s \rangle + \delta E_s$. In the regime of weak-scattering characterized by the condition $|\delta E_s| \ll |\langle E \rangle|$, multiple scattering in the background scattering medium can be described in the mean-field approximation. This approximation is based on well-established result in multiple scattering theory~\cite{sheng2007introduction}, that are summarized in  Appendix A. In the mean-field approach, the fluctuating part $\delta E_s$ of the scattered field at point $\rg_D$ in the detector plane, and at frequency $\omega$, reads
\begin{equation}
\delta E_s(\rg_D) = k_0^2 \int \langle G(\rg_D,\rg^\prime) \rangle \Delta\varepsilon(\rg^{\prime}) \langle E(\rg^\prime) \rangle d^3 r^\prime \, ,
\label{eq:delta_E_mean_field}
\end{equation}
where $k_0=\omega/c$ is the wavenumber in free space, $\langle G(\rg_D,\rg^\prime) \rangle$ is the average Green's function connecting a point $\rg^\prime$ in the sample to the detection point $\rg_D$, and the integration is over the volume of the sample. Here the average Green's function $\langle G \rangle$ and the average field $\langle E \rangle$ are defined as solutions of a propagation equation in an effective medium, described by an effective dielectric function $\epseff$, representing the average contribution of the random scattering background (the detailed derivation of Eq.~(\ref{eq:delta_E_mean_field}) is given in Appendix A). It is important to note that this effective medium description is a rigorous result of multiple scattering theory~\cite{sheng2007introduction,akkermans2007mesoscopic}.
Noting that $\Delta\varepsilon(\rg^{\prime})=\eps(\rg^\prime)-\epseff$ is the local dielectric contrast between a heterogeneity in the sample and the effective medium, a clear physical meaning can be given to Eq.~(\ref{eq:delta_E_mean_field}): It expresses the fluctuating part $\delta E_s$ of the scattered field as the result of single scattering in the effective medium, a result valid in the weak-scattering regime $|\delta E_s| \ll |\langle E \rangle|$. The mean-field approach differs from the Born approximation by the renormalization of the background medium into an effective medium, that accounts in particular for the decay of the average Green's function due to scattering, as we shall see. The average field is connected to the incident field $E_0$ in the source plane by the relation
\begin{equation}
\langle E(\rg) \rangle = A  \int  \langle G(\rg,\rg_S) \rangle E_{0}(\rg_{S}) \, d^{2} \rho_S
\label{eq:average_field}
\end{equation}
where $\langle G(\rg,\rg_S) \rangle$ is the average Green's function connecting a point $\rg_S=(\rhog_S,z_S)$ in the source plane to an arbitrary point $\rg$, $A$ is a constant that we do not need to specify, and the integration is along the source plane. Inserting Eq.~(\ref{eq:average_field}) into Eq.~(\ref{eq:delta_E_mean_field}) leads to
\begin{equation}
\delta E_s(\rg_D) = A \, k_{0}^{2} \int d^{3} r^{\prime} \int d^{2} \rho_S \, \langle G_T(\rg_D,\rg^{\prime}) \rangle \Delta\varepsilon(\rg^{\prime}) \langle G_T(\rg^{\prime},\rg_S) \rangle E_{0}(\rg_{S}) \, .
\label{eq:fluct_scattered_field}
\end{equation}
In this expression, for the sake of clarity, we have denoted by $\langle G_T \rangle$ the Green's function that accounts for propagation through the optics in the sample arm, transmission at the effective medium surface, and propagation inside the effective medium. The average scattered field at the detector is readily deduced from Eq.~(\ref{eq:average_field}), and reads
\begin{equation}
\langle E_s(\rg_D) \rangle = A  \int  \langle G_R(\rg_D,\rg_S) \rangle E_{0}(\rg_{S}) \, d^{2} \rho_S \, .
\label{eq:average_scattered_field}
\end{equation}
Here, the average Green's function $\langle G_R \rangle$ accounts for propagation through the optics in the sample arm, and reflection at the effective medium surface. For practical calculations, the average Green's function in transmission or reflection can be approximated using a simple model, as we shall see below.

\section{FF-OCT signal}
\label{signal_paraxial}

\begin{figure}[h]
	\centering
	\includegraphics[width=9cm]{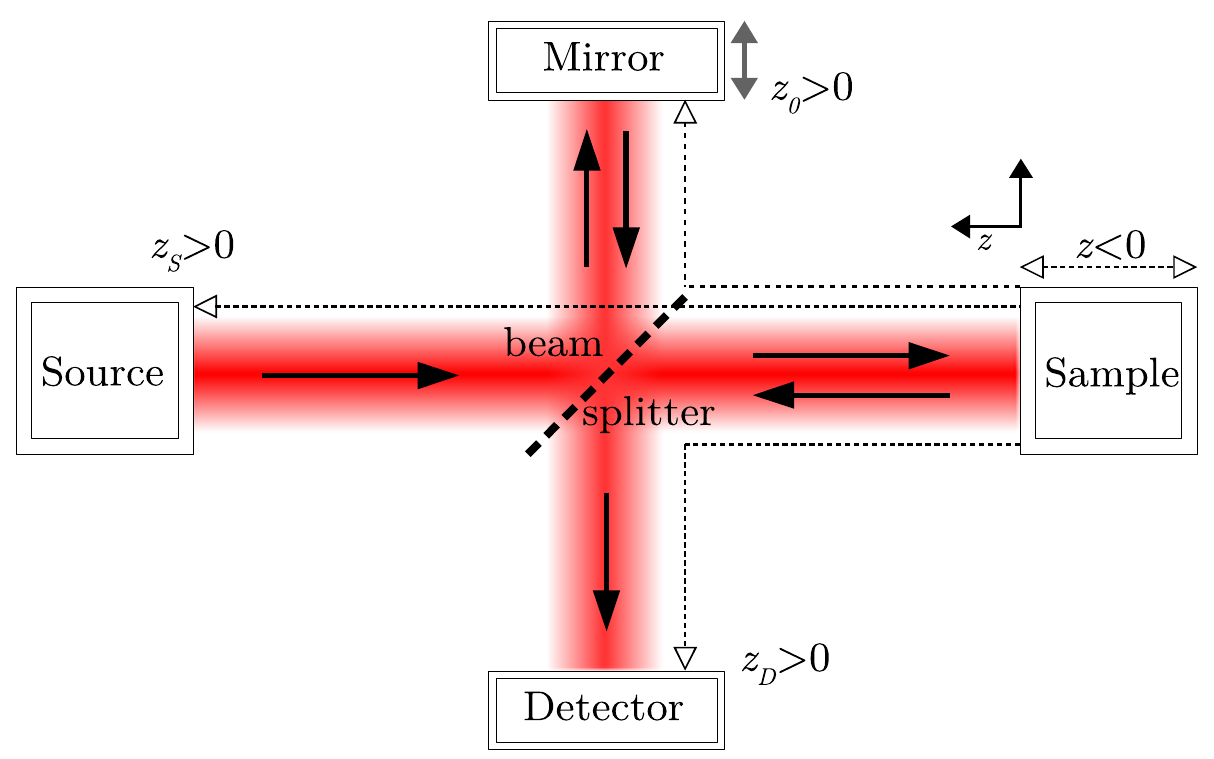}
	\caption{Schematic view of a FF-OCT setup using a source with partial temporal and spatial coherence. The detector measures the interferogram 
between the light backscattered from the sample, and the light reflected on the mirror. Illumination and collection optics (such as microscope objectives)
in the different arms of the interferometer are not represented, since the model does not rely on a particular design. The $z$-axis of the reference frame is chosen so that the plane $z=0$ coincides with the input surface of the sample (referred to as sample plane), with $z<0$ inside the sample. The $z$-axis follows the optical axis in each arm, with the source, reference mirror and detector planes corresponding to $z=z_S$, $z=z_0$ and $z=z_D$, respectively.}
	\label{fig:OCT_setup}
\end{figure}

An OCT setup is based on a Michelson interferometer, as represented schematically in Fig.~\ref{fig:OCT_setup}. Starting from the source, the beam is divided by a beam-splitter to travel along two arms. The sample arm collects the field backscattered from the sample. In the second arm, a reference beam is produced by reflection on a mirror. The detector collects the intensity resulting from the interference between the sample and reference beams. In FF-OCT, a full-field illumination is used in combination with an array of detectors (in practice a CCD camera) to record the signal at multiple transverse locations in parallel, and produce an {\it en face} image. This means that both arms contain a microscope objective, not represented in Fig.~\ref{fig:OCT_setup} for simplicity. The state of coherence of the light source plays a crucial role in FF-OCT. Beyond the longitudinal sectioning given by the temporal coherence length (or equivalently the spectral bandwidth), spatial coherence influences the lateral resolution, as well as the sensitivity to aberrations in the sample arm~\cite{xiao2016full}. In order to understand precisely the role of temporal and spatial coherence, we need to build a model of the FF-OCT signal that includes the scattering model introduced in the previous section, and the main features of the interferometric and broadband detection used in OCT.

Under illumination by statistically stationary light, the intensity recorded by a Michelson interferometer is characterized by three terms: The (time) average intensity of the reference field, the average intensity of the scattered field, and the cross-correlation between them (interference term). Assuming $|E_s| \ll |E_0|$, the signal carrying information on the sample is the interference term (the intensity of the scattered field is negligible, and the intensity of the reference field only contributes as a background signal)~\cite{davis2007autocorrelation}. In the frequency domain, the OCT signal measured at a given point $\rg_D$ in the detector plane is
\begin{equation}
       S(\rg_D, \omega) =  \overline{\Eref^*(\rg_D)  E_{s}( \rg_D) } \, ,
\label{eq:signal_general}
\end{equation}
where $\Eref$ and $E_{s}$ are, respectively, the complex amplitude of the reference and backscattered field at frequency $\omega$, the superscript $*$ denotes the complex conjugate, and the overline means a time averaging over the fluctuations of the partially coherent source. 

The incident field $E_0$ at a point $\rg=(\rhog,z)$ can be written in the form of a plane-wave expansion
\begin{align}
E_0 (\rg) = \int d^2 q  \, e_0 (\qg) \exp [i \qg \cdot \rhog - i k_z(q) z] \, ,
\label{eq:plane_wave_incident}
\end{align}
with $k_z(q) = (k_0^2 - q^2)^{1/2}$. The $z$-axis is chosen such that the plane $z=0$ coincides with the sample plane (the sample filling the half-space $z<0$), and the axis $z>0$ follows the optical axis in each arm (see Fig.\ref{fig:sample}(d)). Assuming a weakly focused beam (the influence the numerical aperture is discussed in section~\ref{sect_NA}), we can perform a zeroth-order paraxial approximation $k_{z}(q) \simeq k_{0}$, resulting in $E_0 (\rg) =  E_{0}(\rhog, z=0)  \exp(-i k_0 z)$. The field in the reference arm is assumed to coincide with the incident field longitudinally shifted by the mirror displacement. Under the same paraxial approximation, we can write $\Eref^*(\rg_D) =  E^{*}_{0}(\rhog_D,z=0)  \exp [-i k_{0} (z_{D}- 2z_{0}) ]$
where $z_0$ is the position of the mirror along the $z$-axis in the reference arm and $z_D$ is the position of the detection plane along the $z$-axis in the detection arm. From Eqs.~(\ref{eq:fluct_scattered_field}), (\ref{eq:average_scattered_field}), (\ref{eq:signal_general}) and the expressions of $E_0$ and $\Eref^*$ above , we can write the FF-OCT signal as
\begin{equation}
 S(\rg_D,\omega) =  \langle S (\rg_D,\omega) \rangle +  \delta S(\rg_D,\omega)
\end{equation}
with
\begin{eqnarray}
\langle S (\rg_D,\omega) \rangle &=& A \exp [i k_{0} (2z_0 - z_{S} - z_{D}) ] \int d^{2} \rho_S \, \langle G_R(\rg_D,\rg_S) \rangle  W_{0}({\rhog}_{S}-{\rhog}_{D}) \, , \label{eq:OCT_signal_average}\\
\delta S(\rg_D,\omega) &=& A k_0^2 \exp [i k_{0} (2z_0 - z_{S} - z_{D}) ]
 \int d^{3}r^{\prime} \int d^{2} \rho_S \, \langle G_T(\rg_D,{\bf r^{\prime}}) \rangle  \Delta\varepsilon(\rg^{\prime}) \nonumber \\
 && \times \, \langle G_T({\bf r^{\prime}},{\rg}_S) \rangle W_{0}({\rhog}_{S}-{\rhog}_{D}) \, . \label{eq:OCT_signal_real}
\end{eqnarray}
In these expressions we have introduced the cross-spectral density
\begin{equation}
W_{0}(\rhog_S-\rhog_D) = \overline{E^{*}_{0}(\rhog_D,z=0) E_{0}(\rhog_S,z=0)} \, ,
\end{equation}
that characterizes the state of coherence of the incident field in the sample plane $z=0$~\cite{mandel1995optical}. Here we assume that the incoming intensity is uniform over the sample surface, and use a homogeneous Schell model such that $W_{0}$ is a function of $\rhog_S - \rhog_D$ only~\cite{marks2009partially}.

It is clear that $\langle S (\rg_D,\omega) \rangle$ characterizes the effective medium, and does not carry information on $\Delta\varepsilon(\rg^{\prime})$. In order to build the simplest model, we assume weak scattering, meaning that $k_0\lext \gg 1$ with $\lext$ the extinction mean free path characterizing the decay of the average field (the exinction coefficient $\mu_e = 1/\lext$ can also be used equivalently). The effective medium, as seen by the average field, is described by a dielectric function  $\epseff = \varepsilon_b + i/(k_0 \lext)$ (this expression being valid to first order in the small parameter $1/(k_0\lext)$)~\cite{sheng2007introduction,akkermans2007mesoscopic}. In this limit, the reflected average Green's function $\langle G_R \rangle$ simply describes reflection at the surface of the effective medium, and contributes as a uniform background. Assuming a low index mismatch with $\varepsilon_b \simeq 1$, the contribution of $\langle S (\rg_D,\omega) \rangle$ becomes negligible. In the following we focus on the contribution $\delta S(\rg_D,\omega)$ that carry the relevant information on the image formation process. It is useful to introduce the plane-wave expansion of the  average Green's function: 
\begin{align}
\langle G_T(\rg,\rg^\prime) \rangle =
\frac{i}{2\pi}\int d^{2}q \frac{g({\bf q})}{k_{z}(q)} \exp \big[ i {\bf{q}} \cdot ( \rhog - \rhog^{\prime}) +  i k_{z}(q) \lvert z-z^{\prime}\rvert \big] \, .
\label{eq:G_plane_wave}
\end{align}
We take $k_z(q)=(k_0^2-q^2)^{1/2}$ for propagation outside the sample (for $z,z^\prime >0$), and $k_z(q)=\kzeff (q)=(\epseff k_0^2-q^2)^{1/2}$ for propagation inside the sample (for $z,z^\prime <0$). The filter $g(\qg)$ plays the role of a pupil function that limits the transverse wavevector $\qg$ within a region bounded by the numerial aperture $NA$ of the objectives (the simplest model for $g(\qg)$ is a disk with radius $k_0 \, NA$)~\cite{sentenac2018tutorial}. For $g(\qg)=1$ (infinite pupil), we recover the Weyl expansion of the free-space Green's function. To get an expression of the FF-OCT signal relevant for an analysis in terms of optical transfer function, we insert Eq.~(\ref{eq:G_plane_wave}) into Eq.~(\ref{eq:OCT_signal_real}), and perform again a zeroth-order paraxial approximation, using $k_z(q) \simeq k_0$ outside the sample, and $\kzeff(q) \simeq k_0 + i/(2\lext)$ inside the sample. The result is easily written in terms of the 2D Fourier transform of the signal, 
$\widetilde{\delta S}( {\bf q},\omega) = \int \delta S(\rg_D,\omega) \exp(-i\qg \cdot \rhog_D) d^2 \rho_D/{4\pi^2}$, 
and reads
\begin{align}
\widetilde{\delta S}( {\bf q},\omega) = - 4\pi^2 A \int d z^\prime 
\exp [2i k_{0} (z_{0}-z^\prime) ] \, \widetilde{\Delta\varepsilon}({\bf q},z^\prime) \exp(-\lvert z^{\prime} \rvert/\lext) \nonumber \\
 \times \int d^2q^\prime g({\bf q}^\prime)  g({\bf q}+{\bf q}^\prime) \widetilde{W}_0({\bf q}^\prime) \, ,
 \label{eq:OCT_signal_2}
\end{align}
where $\widetilde{\Delta\varepsilon}({\bf q},z^\prime)$ is the 2D Fourier transform of $\Delta\varepsilon(\rhog^\prime,z^\prime)$, and $\widetilde{W}_0({\bf q}^\prime)$ is the 2D Fourier transform of $W_{0}({\rhog}_{S}-{\rhog}_{D})$. Note that $\widetilde{W}_0({\bf q}^{\prime}) \propto {I}_0({\bf q}^{\prime})$, with $I_0({\bf q}^{\prime})$ the angular distribution of the incident intensity~\cite{barabanenkov1968radiation,mandel1995optical}. Also note that this expression does not rely on any assumption regarding the optics in the illumination, reference and detection arms, and does not depend on the definition of a focal plane for either illumination of detection (these features drive the precise form of the Green's function $g({\bf q})$). In practice, in order to control the degree of spatial coherence in the sample plane $z=0$, one could choose to control the intensity distribution in the source plane $z=z_S$ and conjugate this plane with the plane $z=0$, but this practical choice does not influence the general form of Eq.~(\ref{eq:OCT_signal_2}).

A feature of OCT is to integrate the signal over a broad spectral range. For a statistically stationary source, the broadband signal is obtained by integrating the different frequency components over the source bandwidth $\Delta \omega$. Assuming that $W_0$ and $\Delta\varepsilon$ have a weak dependence on $\omega$, the spectral integration gives 
\begin{equation}
             \int^{\omega_{0} + \Delta \omega /2}_{\omega_0- \Delta \omega /2}
              \exp [ 2 i k_0 (z_{0} - z^{\prime}) ] d\omega = \  \exp [ 2 i \bar{k}_0 (z_{0} - z^{\prime}) ] \, \mathrm{sinc}[(z_0 - z^\prime)\Delta \omega/c] \, \Delta\omega \; ,
\label{eq:int_omega_sinc}
\end{equation}
where $\bar{k}_0=\omega_0/c$, with $\omega_0$ the central frequency of the source.
The $\mathrm{sinc}$ function, considered as a function of $z^\prime$, is centered at $z_0$ with a width $\ell_\omega=c/\Delta\omega$ corresponding to the temporal coherence length of the incident light, and is responsible for the longitudinal sectioning in OCT. When $\ell_{\omega} \ll \lext $, we can use the approximation
$\mathrm{sinc}[(z_0 - z^\prime)/\ell_\omega] \simeq \pi \ell_\omega \delta(z_0 - z^\prime)$. We endup with a closed-form expression of the 2D Fourier transform of the broadband OCT signal $\widetilde{\delta S}( {\bf q})=\int_{\Delta\omega} \widetilde{\delta S}( {\bf q},\omega) d\omega$, that depends on the position $z_0$ of the mirror and on the source bandwith $\Delta\omega$: 
\begin{align}
\label{Final_S}
\widetilde{\delta S}( {\bf q})  =  - 4 \pi^3 A c \
 \widetilde{\Delta\varepsilon}({\bf q},z_0) \exp(- z_0/ \lext) 
\int d^2q^\prime g({\bf q}^\prime) g({\bf q}+{\bf q}^\prime)  \widetilde{W}_0({\bf q}^\prime) \, .
\end{align}
Expression (\ref{Final_S}) of the the FF-OCT signal implicitly involves different length scales, whose interplay is critical in the analysis of the signal:  
The length scale $\ell_\varepsilon$ characterizing the microscopic random inhomogeneities in the sample, the length scale $L_\varepsilon$
characterizing large-scale deterministic fluctuations of the dielectric function, the temporal coherence length $\ell_{\omega}$ and the spatial 
coherence length $\ell_{c}$ of the incident light. Expression (\ref{Final_S}) is similar to that previously derived in Ref.~\cite{marks2009partially}. 
The main difference is that our derivation includes a scattering model based on a mean-field approach, that only relies on the weak-scattering assumption
$|\delta E_s| \ll |\langle E \rangle|$, always satisfied when the scattered field remains much smaller than the incident field. This approach goes beyond a first-order 
or second-order Born approximation, by accounting rigorously for the propagation of the average field in a renormalized effective medium.
An interesting consequence is that the extinction of the signal with depth, due to scattering and absorption, emerges explicitly.

\section{Spatial coherence and response function }

Expression (\ref{Final_S}) is an interesting starting point for the analysis of the role of spatial coherence on the image formation in FF-OCT. To proceed, let us rewrite it in the compact form
\begin{align}
\label{Signal_transfer}
\widetilde{\delta S}( {\bf q})  =  \widetilde{H}(\qg) \, \widetilde{\Delta\varepsilon}_{z_0}({\bf q}) \, ,
\end{align}
where $\widetilde{H}(\qg) \sim \int d^2q^\prime g({\bf q}^\prime) g({\bf q}+{\bf q}^\prime)  \widetilde{W}_0({\bf q}^\prime)$ is the FF-OCT transfer function (we omit constant prefactors for simplicity), and $\widetilde{\Delta\varepsilon}_{z_0}({\bf q}) = \widetilde{\Delta\varepsilon}({\bf q},z_0) \exp(- z_0/ \lext)$ is the weighted 2D Fourier transform of the dielectric contrast. The width of the cross-spectral density $W_0(\rhog)$ defines the spatial coherence length $\ell_c$ of the incident field in the sample plane $z=0$. When $\ell_c \gg L_\varepsilon \gg \ell_\varepsilon$, the incident light can be considered fully coherent, which corresponds to $\widetilde{W}_0({\bf q}^\prime) \sim I_0 \delta(\qg^\prime)$, $I_0$ being proportionnal to the incident intensity. The transfer function for spatially coherent illumination becomes
\begin{align}
\label{H_coherent}
 \widetilde{H}_c(\qg)  \sim  I_0 \, g(0) g({\bf q}) \, .
\end{align}
The coherent transfer function $\widetilde{H}_c(\qg)$ posseses the same spatial frequency content as $g(\qg)$, and covers the spatial frequency range $q \leq \bar{k}_0 NA$, leading to a transverse spatial resolution limit $\lambda/(2 NA)$, where $\lambda=2\pi/\bar{k}_0$ is the central wavelength of the incident field. The regime of spatially incoherent illumination corresponds formally to $W_0(\rhog) \sim \delta(\rhog)$, or equivalently $\widetilde{W}_0({\bf q}^\prime) = I_0^\prime$. The transfer function for spatially incoherent illumination reads as
\begin{align}
\label{H_incoherent}
 \widetilde{H}_i(\qg)  \sim  I_0^\prime \, \int d^2q^\prime g({\bf q}^\prime) g({\bf q}+{\bf q}^\prime) \, .
 \end{align}
In practice, since $\ell_c$ cannot be made smaller than $\lambda/2$ under far-field illumination, the condition $\ell_c \ll \ell_\varepsilon \ll L_\varepsilon$ that rigorously corresponds to  $W_0(\rhog) \sim \delta(\rhog)$ is out of reach. Nevertheless, provided that FF-OCT is used to image the large-scale inhomonegeities at the scale $L_\varepsilon$, and not to resolve the small-scale random scattering centers, the condition of incoherent illumination can be relaxed to be $\ell_c \simeq \ell_\varepsilon \ll L_\varepsilon$.
The integral term in Eq.~(\ref{H_incoherent}) shows that the spatial frequency range encompassed by $\widetilde{H}_i(\qg)$ is broader by a factor of two compared to coherent illumination, which is a usual result in optical microscopy, leading to a theoretical resolution limit $\lambda/(4 NA)$.
The results above, for coherent and incoherent illumination, are identical to those previsouly established in Ref.~\cite{marks2009partially}, and in agreement with the qualitative analysis of coherent and incoherent FF-OCT presented in Ref.~\cite{sentenac2018tutorial}. 

Expressions (\ref{H_coherent}) and (\ref{H_incoherent}) may also provide a theoretical frame to study the influence of aberrations in the sample arm on the image quality, recently discussed in Ref.~\cite{xiao2016full}. A precise study is beyond the scope of the present work. Qualitatively, we can simply note that weak aberrations that would tend to squeeze the Green's function $g({\bf q})$ in Fourier space (without changing the cutoff frequency) would narrow the shape of $\widetilde{H}_c(\qg)$, thus degrading the image quality. For incoherent illumination, the convolution product in Eq.~(\ref{H_incoherent}) reduces the sensitivity of the transfer function to change in shapes due to weak aberrations, resulting in a better protection of the image quality.

\section{Depth dependence of the spatially averaged signal}
\label{sect_NA}

The decay of the signal with depth is a feature of OCT. For paraxial illumination and detection, and in a medium with 
$\widetilde{\Delta\varepsilon}({\bf q},z_0)$ independent of $z_0$, the signal follows an exponential decay $\exp(-z_0/\lext)$, as described by Eq.~(\ref{Final_S}). 
This exponential decay can be used for the measurement of the extinction length in weakly scattering materials~\cite{plamann2019OCT}.
At higher numerical aperture, a correction to a pure exponential decay is expected, that may also depend on the degree of spatial coherence of the incident field. To address this question, we make use of a second-order paraxial approximation
\begin{eqnarray}
k_z(q) &\simeq& k_0 - \frac{1}{2} \frac{q^2}{k_0} \, , \label{eq:para2_1} \\
\kzeff (q) &\simeq& \sqrt{\epseff } k_0 - \frac{1}{2} \frac{q^2}{\sqrt{\epseff} k_0} \simeq k_0 +\frac{i}{2\lext} - \frac{1}{2} \frac{q^2}{ k_0} + \frac{i q^2}{4k_0^2 \lext} \, , \label{eq:para2_2}
\end{eqnarray}
in the plane-wave expansions of the incident field [Eq.~(\ref{eq:plane_wave_incident})] and of the average Green's function [Eq.~(\ref{eq:G_plane_wave})]. Note that, as a result of the mean-field approach,
the paraxial approximation in the scattering medium amounts to replacing $k_0$ by $\sqrt{\epseff } k_0$. This strongly influences the dependence of the FF-OCT on the numerical aperture, as we shall see.
Following again the steps leading to Eq.~(\ref{Final_S}), we end up with an expression of the FF-OCT signal that explicitely accounts for the angular aperture of the illumination and detection beams. In order to discuss the depth dependence of the spatially integrated signal, we consider $\widetilde{\delta S}( {\bf q}=0)$, which can be cast in the following form (details of the derivation are given in Appendix B):
\begin{align}
\label{Final_S_NA}
\widetilde{\delta S}( {\bf q}=0)  = - 4\pi^3 A c \, \widetilde{\Delta\varepsilon}({\bf q}=0,z_0) \exp(-z_0/\lext) 
\int d^2q^\prime g^2({\bf q}^\prime) \widetilde{W}_0({\bf q}^\prime) \exp[-q'^2 z_0/(2\bar{k}_0^2\lext)] \, ,
 \end{align}
 where $\bar{k}_0= \omega_0/c$ is the central wavenumber of the polychromatic incident field.
This expression is well suited to a discussion of the depth dependence of the integrated FF-OCT signal. Note that in practice, an exponential fit to the signal is often used to estimate the extinction mean free path $\lext$ in statistically homogeneous and isotropic scattering media. From Eq.~(\ref{Final_S_NA}), it is clear that at finite $NA$ a deviation from a pure exponential decay $\exp(-z_0/\lext)$ may be observed due to the dependence on $z_0$ of the integral over $q'$. This integral contains several cutoffs. First, the last exponential term gives a depth-dependent cutoff $q'_{z_0} \simeq \bar{k}_0 (\lext/z_0)^{1/2}$. For $z_0 \gg \lext$, the integral is restricted to vanishingly small $q'$, and a pure exponential decay $\exp(-z_0/\lext)$ is always expected in the tail of the signal versus depth. Second, the angular spectrum of the Green's function $g(\qg')$ produces a cutoff $q'_{NA} \simeq \bar{k}_0NA$, showing that for $NA \to 0$ a pure exponential decay is observed, in agreement with Eq.~(\ref{Final_S}). Third, the Fourier transform of the cross spectral density $\widetilde{W}_0({\bf q}^\prime)$ introduces a cutoff $q'_c \simeq 2\pi/\ell_c$ that depends on the degree of spatial coherence of the incident light in the sample plane. For $\ell_c \to \infty$ (spatially coherent illumination), a pure exponential decay is also observed. More precisely, to prevent the $\exp[-q'^2 z_0/(2\bar{k}_0^2\lext)]$ term in the integral to play a role, we need either $NA \ll (\lext/z_0)^{1/2}$ (low numerical aperture regime), or $\ell_c \gg \lambda (z_0/\lext)^{1/2}$ (coherent illumination).

\begin{figure}[h]
	\centering
	  \includegraphics[width=7.5cm]{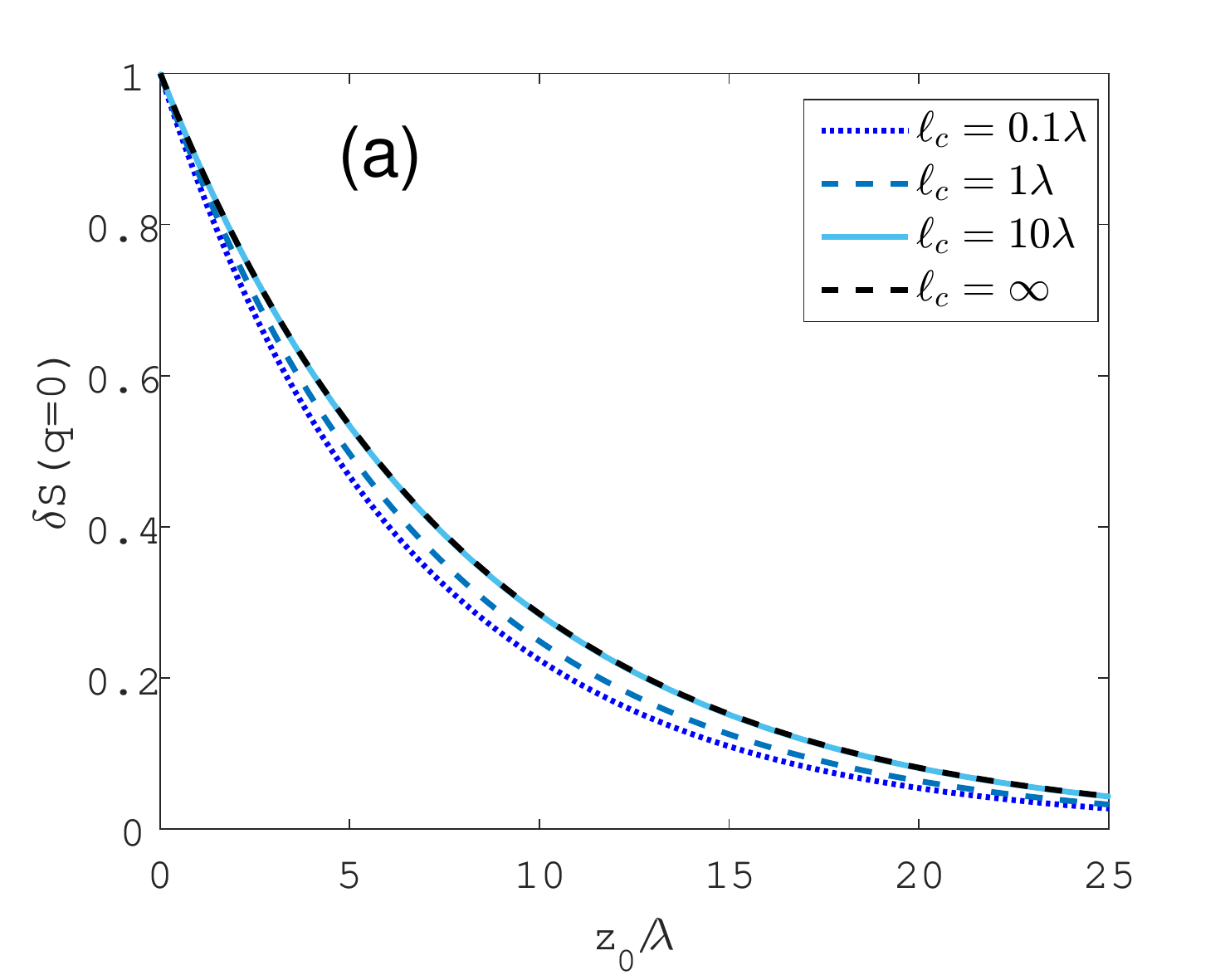}
      \includegraphics[width=7.5cm]{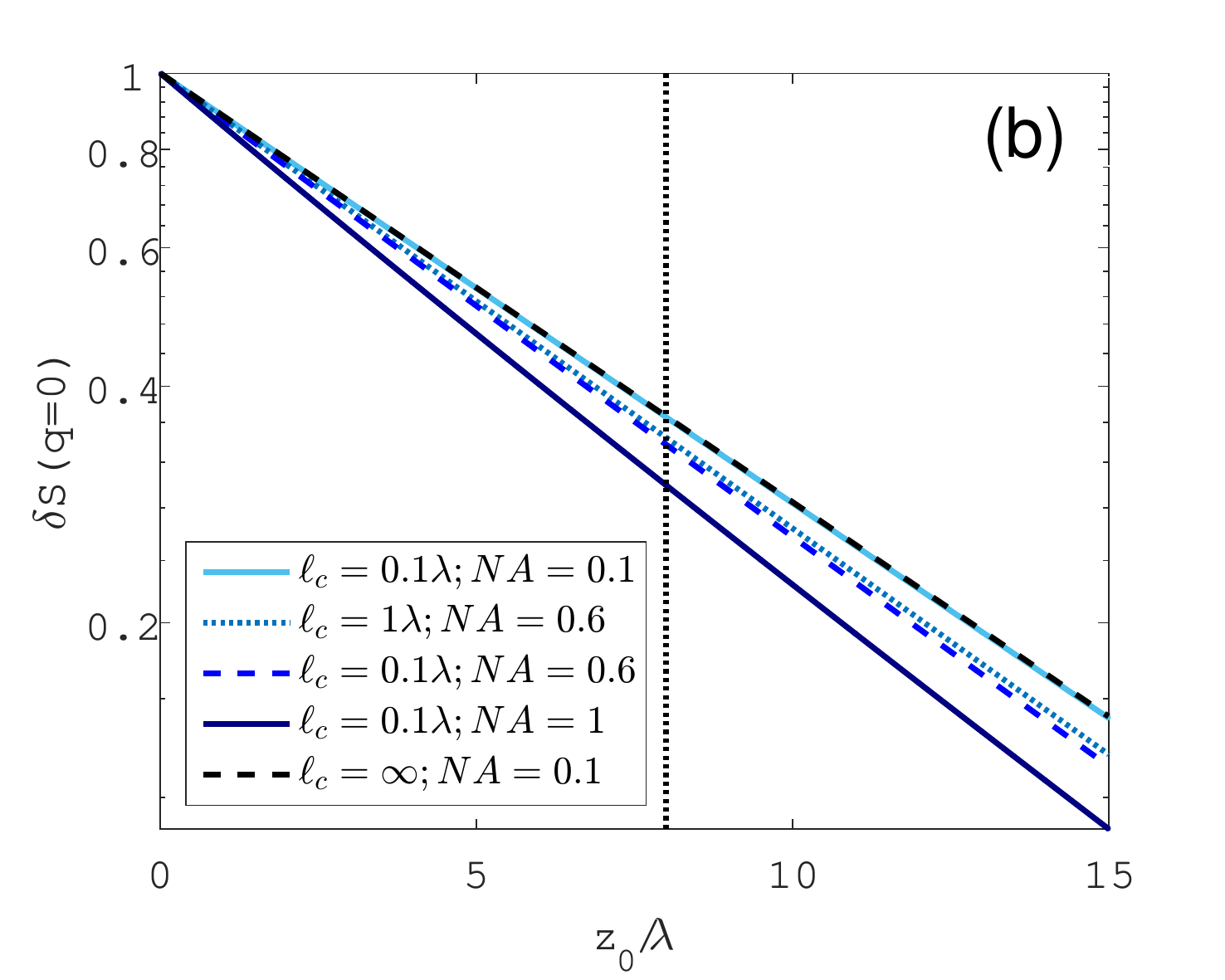}
	\caption{Spatially integrated FF-OCT signal $\widetilde{\delta S}( {\bf q}=0)$ versus the normalized depth $z_0/\lambda$ for different values of the spatial coherence length $\ell_c$ and numerical apertures $NA$. The curves are normalized by their value at $z_0=0$. The central wavelength of the incident light is $\lambda = 800$ nm, the temporal coherence length is $\ell_\omega = 1$ $\mu$m, and the extinction length is $\lext = 8\lambda$. The pupil function $g({\bf q})$ is modeled by a Gaussian profile $g({\bf q})\propto\exp[-q^2/ (\bar{k}_0NA)^2 ]$. The cross-spectral density $\widetilde{W}_0({\bf q})$ in the Gaussian Shell-model is $\widetilde{W}_0({\bf q})\propto\exp[-q^2\ell_c^2/(4\pi^2)]$. (a): Signal for different degrees of spatial coherence with $NA=1$. (b): Signal for different values of $\ell_c$ and $NA$. The vertical dashed line corresponds to $z_0=\lext$.}
	\label{fig:decay}
\end{figure}

This qualitative analysis is confirmed by numerical calculations of the spatially integrated signal using Eq.~(\ref{Final_S_NA}), as shown in Fig.~\ref{fig:decay}. A non-exponential decay is observed in the regime $NA \simeq  (\lext/z_0)^{1/2} \simeq \lambda/\ell_c$. The deviation from a pure exponential decay is enhanced at high $NA$ and low spatial coherence. Although the curves are displayed in the regime $z_0 \simeq \lext$, we have verified that an exponential decay $\exp(-z_0/\lext)$ is recovered in any case when $z_0 \gg \lext$. Numerical calculations based on {\it ab initio} simulations could extend the analysis beyond the weak-scattering regime and second-order paraxial approximations used in this model. This is left for future work.

\section{Conclusion}

In summary, we have developed a model of FF-OCT that includes a mean-field scattering theory, in addition to a precise description of temporal and spatial coherence. The model describes the decay of the signal with depth due to scattering and absorption, an essential feature in OCT imaging. It also permits an analysis of the interplay between different length scales characterizing the scattering medium and the degree of coherence of the incident field. Based on explicit expressions of the FF-OCT signal, we have discussed several features of FF-OCT imaging. We have analyzed the influence of spatial coherence on the optical transfer function, and discussed the particular cases of fully coherent and incoherent illuminations. We have also studied the depth dependence of the signal integrated over the transverse directions, that not only limits the penetration depth in OCT, but is also used to measure the extinction length in scattering materials. For spatially incoherent illumination, and/or with high numerical aperture of the illumination/detection optics, deviations from a pure exponential decay $\exp(-z_0/\lext)$ can be observed. A pure exponential decay is always recovered in the tail of the signal versus depth when $z_0 \gg \lext$. Our analysis provides a clear frame for the use of the Beer-Lambert law for quantitative measurements of the extinction length in scattering media. The model also provides a framework for a precise analysis of the role of aberrations generated by the optics in the sample arm, or by the scattering medum itself, and for the development of advanced inverse reconstruction procedures going beyond the first and second-order Born approximations. 

\section*{Funding}
This work was supported by the European Union's Seventh Framework Program under HELMHOLTZ grant agreement No 610110, and by LABEX WIFI (Laboratory of Excellence ANR-10-LABX-24) within the French Program ``Investments for the Future'' under reference ANR-10-IDEX-0001-02 PSL$^{\ast}$.

\section*{Acknowledgments}

We thank A. Aubry, V. Barolle, C. Boccara, R. Bocheux, K. Irsch and K. Plamann for stimulating discussions.

\appendix

\section{Elements of multiple scattering theory}

In order to build a mean-field expression of the scattered field, we first consider a purely scattering medium with 
dielectric function $\eps_b + \delta\eps(\rg)$, as represented in Fig.~\ref{fig:sample}(b). We assume that the random
variable $\delta\eps(\rg)$, that describes a spatial distribution of small-scale scattering centers, is
statistically homogeneous and isotropic. An important result of multiple scattering theory, states that the 
average field obeys a propagation equation in an effective homogeneous medium, as represented in Fig.~\ref{fig:sample}(c). 
We take this result as our starting point (for a derivation, see for example Refs.~\cite{sheng2007introduction,akkermans2007mesoscopic}). 
More precisely, the average field obeys
\begin{equation}
\nabla^2 \langle E(\rg) \rangle + k_0^2 \epseff  \, \langle E(\rg) \rangle  = 0 \, ,
\label{eq:Dyson_E}
\end{equation}
where $\epseff$ is the effective dielectric function that describes the average contributio of the random scattering medium.
Equivalently, the average Green's function satisfies the Dyson equation
\begin{equation}
\nabla^2 \langle G(\rg,\rg^\prime) \rangle + k_0^2 \epseff  \, \langle G(\rg,\rg^\prime) \rangle  = -\delta(\rg-\rg^\prime) \, ,
\label{eq:Dyson_G}
\end{equation}
with an outgoing wave condition at infinity.
This effective medium approach is strictly valid provided that $\lcor \ll \lambda$, meaning that non-locality in the effective dielectric 
function can be neglected, $\lcor$ being the microscopic clength scale associted to $\delta\eps(\rg)$ \cite{sheng2007introduction}.
Interestingly, the imaginary part of the effective dielectric constant describes the attenuation of the average field by scattering (and absorption), and defines the 
extinction mean free path $\lext$ such that $\mathrm{Im} \epseff  = (k_0\lext)^{-1}$. In absence of absorption, the extinction mean free path
coincides with the scattering mean free path~$\lsca$.

In the general situation represented in Fig.~\ref{fig:sample}(a), in which large-scale deterministic objects are superimposed to the random scattering
background, the dielectric function is $\eps(\rg) = \eps_b + \eps_d(\rg)+ \delta\eps(\rg)$. The total field obeys
\begin{equation}
\nabla^2 E(\rg) + k_0^2 \eps(\rg) E(\rg)   = 0 \; .
\label{eq:total_E}
\end{equation}
In order to obtain the equation satisfied by the fluctuating scattered field $\delta E_s = E-\langle E \rangle$, we subtract Eq.~(\ref{eq:Dyson_E}) to Eq.~(\ref{eq:total_E}), which leads to
\begin{equation}
\nabla^2 \delta E_s(\rg) + k_0^2 \epseff  \, \delta E_s(\rg)  = -k_0^2 [\eps(\rg)-\epseff ] E(\rg) \; .
\label{eq:delta_E}
\end{equation}
Using the average Green's function $\langle G(\rg,\rg^\prime) \rangle$ defined in Eq.~(\ref{eq:Dyson_G}), the solution
to Eq.~(\ref{eq:delta_E}) is shown to obey the following integral equation:
\begin{equation}
\delta E_s(\rg) = k_0^2 \int \langle G(\rg,\rg^\prime) \rangle [\eps(\rg^\prime)-\epseff ] E(\rg^\prime) d^3 r^\prime \; .
\label{eq:delta_E_integral}
\end{equation}
Assuming that scattering is sufficiently weak for the condition $|\delta E_s| \ll |\langle E \rangle |$ to be valid, the above expression
can be approximated by
\begin{equation}
\delta E_s(\rg) = k_0^2 \int \langle G(\rg,\rg^\prime) \rangle [\eps(\rg^\prime)-\epseff ] \langle E(\rg^\prime) \rangle d^3 r^\prime \; .
\label{eq:delta_E_mean_field_app}
\end{equation}
Equation (\ref{eq:delta_E_mean_field_app}) is a mean-field approximation to Eq.~(\ref{eq:delta_E_integral}). Physically, it describes
the scattered field in one realization of the medium as resulting from a single scattering process in the effective medium. 

\section{FF-OCT signal in the second-order paraxial approximation}

In this appendix we develop the steps leading to Eq.~(\ref{Final_S_NA}). We start by performing the expansion (\ref{eq:para2_1}) in Eq.~(\ref{eq:plane_wave_incident}), resulting in the following expressions of the incident and reference fields:
\begin{eqnarray}
E_0 (\rg_S) =  \exp (-ik_0 z_S) \int d^2 q \, e_0(\qg) \exp (i \qg \cdot \rhog_S) \exp [iq^2 z_S/(2k_0)] \, , \\
E_{ref}^* (\rg_D) =  \exp [-ik_0 (z_D-2z_0)] \int d^2 q \,  e_0^*(\qg) \exp (-i \qg \cdot \rhog_D) \exp [iq^2 (z_D-2z_0)/(2k_0)] \, . 
\end{eqnarray}
Next we use expansions (\ref{eq:para2_1}) and (\ref{eq:para2_2}) in the expressions (\ref{eq:G_plane_wave}) of the average Green's function. Here the expansion to second order in $q$ is performed only in the exponential term, keeping a zeroth-order expansion $k_z(q) \simeq k_0$ in the denominator that plays a minor role in the overall $z$ dependence. Using the resulting expressions of $E_0$,  $E_{ref}^*$ and $\langle G_T \rangle$ into Eqs.~(\ref{eq:fluct_scattered_field}) and (\ref{eq:signal_general}), we obtain
\begin{align}
\label{eq:app_signal_1}
\widetilde{\delta S}( {\bf q},\omega)  = - 4\pi^2 A \int d z^\prime \exp [2i k_{0} (z_{0}-z^\prime) ] \, \widetilde{\Delta\varepsilon}({\bf q},z^\prime) 
\exp(-\lvert z^{\prime} \rvert/\lext) \nonumber \\
 \times \exp[iq^2 (z'-z_D)/(2k_0)] \exp[-q^2 |z'|/(4k_0^2\lext)]
 \int d^2q^\prime g({\bf q}^\prime)  g({\bf q}+{\bf q}^\prime) \widetilde{W}_0({\bf q}^\prime)  \nonumber \\
 \times \exp[iqq'(z'-z_D)/k_0]  \exp[iq'^2 (z'-z_0)/k_0] \exp[-(qq'+q'^2) |z'|/(2k_0^2\lext)] \, .
\end{align}
This expression extends Eq.~(\ref{eq:OCT_signal_2}) to incident and detection beams with non-negligible numerical apertures.
We now focus on the expression of the signal integrated over the transverse direction, which is obtained by taking ${\bf q}=0$:
\begin{align}
\label{eq:app_signal_2}
\widetilde{\delta S}( {\bf q}=0,\omega)  = - 4\pi^2 A \int d z^\prime \exp [2i k_{0} (z_{0}-z^\prime) ] \, \widetilde{\Delta\varepsilon}({\bf q}=0,z^\prime) 
\exp(-\lvert z^{\prime} \rvert/\lext) \nonumber \\
 \int d^2q^\prime g^2({\bf q}^\prime) \widetilde{W}_0({\bf q}^\prime) \exp[i q'^2 (z'-z_0)/k_0] \exp[-q'^2 |z'|/(2k_0^2\lext)] \, .
 \end{align}
It can be verified by a numerical evaluation that the integral over $q^\prime$ weakly depends on frequency over a bandwidth $\Delta\omega$ corresponding to a depth resolution on the order of one micrometer, which corresponds to $\ell_{\omega} \simeq 1 \mu m$. Making use of Eq.~(\ref{eq:int_omega_sinc}) in the limit $\ell_\omega \ll \lext$, the signal integrated over frequencies becomes
\begin{align}
\label{eq:app_signal_3}
\widetilde{\delta S}( {\bf q}=0)  = - 4\pi^3 A c \, \widetilde{\Delta\varepsilon}({\bf q}=0,z_0) \exp(-z_0/\lext) 
\int d^2q^\prime g^2({\bf q}^\prime) \widetilde{W}_0({\bf q}^\prime) \exp[-q'^2 z_0/(2\bar{k}_0^2\lext)] \, ,
 \end{align}
where $\bar{k}_0= \omega_0/c$ is the central wavenumber of the incident field. This expression is Eq.~(\ref{Final_S_NA}) of the main text. 


\end{document}